\documentclass[
  aps,prd,
  reprint,
  amsmath,amssymb,
  superscriptaddress,
  nofootinbib,
  floatfix
]{revtex4-2}

\usepackage{graphicx}
\usepackage{bm}
\usepackage{xcolor}
\usepackage[colorlinks=true,allcolors=blue]{hyperref}
\hypersetup{
  pdftitle={Perturbative QCD Constraints on Hot Neutron-Star Matter},
  pdfauthor={Tyler Gorda, Aleksi Kurkela, Tuhin Malik}
}

\newcommand{\shat}{\hat{s}}
\newcommand{\nB}{n_\mathrm{B}}
\newcommand{\muB}{\mu_\mathrm{B}}
\renewcommand{\L}{\mathrm{L}}
\renewcommand{\H}{\mathrm{H}}
\newcommand{\nsat}{n_{\rm sat}}
\newcommand{\fmc}{\,\mathrm{fm^{-3}}}
\newcommand{\nstar}{n^{*}}

\makeatletter
\g@addto@macro\bfseries{\boldmath}
\makeatother

\begin{document}

\title{Perturbative QCD Constraints on Hot Neutron-Star Matter}

\author{Tyler Gorda}
\email{gorda.1@osu.edu}
\affiliation{Department of Physics, The Ohio State University, Columbus, OH 43210, USA}
\affiliation{Center for Cosmology and AstroParticle Physics (CCAPP), The Ohio State University, Columbus, OH 43210, USA}

\author{Aleksi Kurkela}
\email{aleksi.kurkela@uis.no}
\affiliation{Faculty of Science and Technology, University of Stavanger, 4036 Stavanger, Norway}

\author{Tuhin Malik}
\email{tuhin.malik@gmail.com}
\affiliation{CFisUC, Department of Physics, University of Coimbra,
  3004-516 Coimbra, Portugal}

\begin{abstract}
Perturbative QCD, supplemented only by thermodynamic consistency, causality, and mechanical stability, constrains the equation of state of dense matter far below the densities at which the perturbative expansion itself converges.
Until now, however, such constraints have been established only for cold, catalyzed matter.
We show that this model-independent bound generalizes to finite temperature along adiabats labelled by the entropy per baryon $\shat$, with the specific enthalpy $h=(\varepsilon+p)/\nB$ assuming the role played by the baryon chemical potential at $T=0$.
We find that increasing the temperature shrinks the allowed region monotonically in a form that is universal across structurally distinct mean-field families.
For DD2, the density at which the equation of state becomes inconsistent with perturbative QCD decreases from $1.0$ to $0.74\fmc$ as $\shat$ increases from $0$ to $5$, falling below the central density of the heaviest cold star.
By $\shat=4$, seven of eleven relativistic mean-field models are excluded at $6\,\nsat$; approximately half of the equations of state drawn from two Bayesian posteriors are excluded by $\shat\simeq3.4$--$4.4$.
When mapped onto the QCD phase diagram, the bound becomes active at temperatures of $100$--$150$~MeV, precisely the regime transiently realized in neutron-star mergers and proto-neutron stars.
\end{abstract}
\date{November 10, 2026}

\maketitle

\section{Introduction}
Perturbative QCD (pQCD) provides a first-principles description of strongly interacting matter at asymptotically high baryon density~\cite{Kurkela:2009gj, Gorda:2021kme}. Although the baryon densities $\nB$ where pQCD is reliable ($\nB\gtrsim40\,\nsat$, with $\nsat \approx 0.16\,\fmc$) lie far above those realized in neutron-star (NS) cores, the pQCD result still matters astrophysically: combined only with thermodynamic consistency, causality ($c_s^2\le1$), and mechanical stability, it constrains the equation of state (EOS) at all lower densities. This is the model-independent construction of Komoltsev and Kurkela~\cite{Komoltsev:2021jzg, Komoltsev:2023zor}, which has become a standard ingredient in NS EOS inference~\cite{Annala:2019puf}.

To date this program has been carried out exclusively at zero temperature $T$, where matter is cold and catalyzed. Yet the environments that carry the most information about dense matter, such as binary NS mergers, core-collapse supernovae, and proto-neutron stars (PNSs), are hot, reaching temperatures of tens to $\sim100~\mathrm{MeV}$ at several times saturation density~\cite{Most:2018eaw,Perego:2019adq,Hanauske:2016gia}. The finite-temperature EOS tables that close the hydrodynamic equations in these simulations are, without exception, built on specific nuclear models~\cite{Oertel:2016bki,Raduta:2021coc}. Whether the pQCD limit says anything model-independent about hot and dense matter has remained an open question, as the bound of \cite{Komoltsev:2021jzg} is formulated and applied only at zero temperature.

In this communication we answer this question. We show that this model-independent bound admits a natural finite-temperature generalization along adiabats of fixed entropy per baryon $\shat\equiv s/\nB$, the variable that labels fluid elements in ideal-hydrodynamic flows. 

\begin{figure*}[t]
  \centering
  \includegraphics[width=\textwidth]{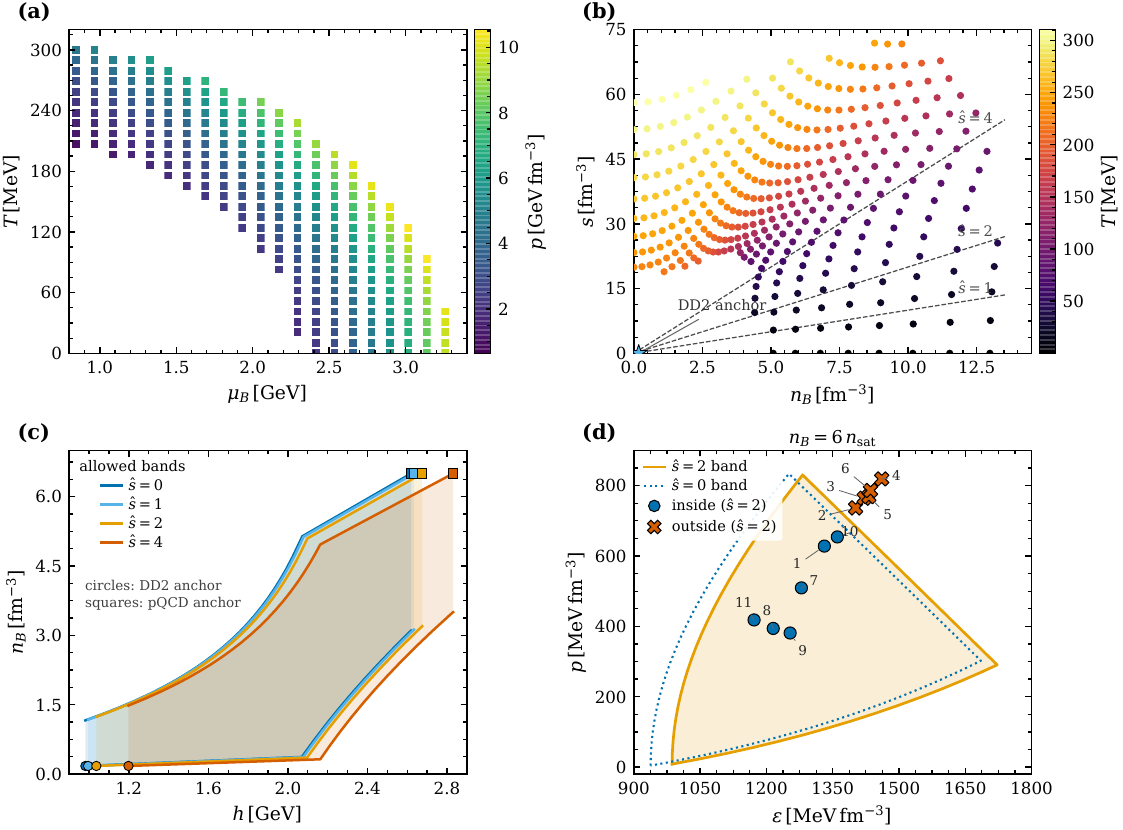}
 \caption{  Overview of the finite-$T$ pQCD input and the model-independent construction. \textbf{(a)} Scale-averaged pQCD pressure on the $(\muB,T)$ plane. \textbf{(b)} The same data mapped onto the $(\nB,s)$ plane and colored by temperature, with radial $\shat=s/\nB$ rays and the low-density DD2 anchor ($\nB=0.176\fmc$, star). \textbf{(c)} Allowed $(h,\nB)$ bands for $\shat=0,1,2,4$; circles and squares mark the DD2 (low) and pQCD (high) anchors. \textbf{(d)} Allowed $(\varepsilon,p)$ region at $6\,\nsat$ for $\shat=2$ (shaded) with eleven RMF EOSs evaluated on their $\shat=2$ isentropes: 1~Char~\cite{Char:2023fue}, 2~DD-ME1~\cite{Niksic:2002yp}, 3~DD-ME2~\cite{Lalazissis:2005de}, 4~DD-MEX~\cite{Taninah:2019cku}, 5~DD-MEX1 and 6~DD-MEX2~\cite{Taninah:2023anchor}, 7~DDB~\cite{Malik:2022zol}, 8~DDV~\cite{Typel:2020ozc}, 9~FSU2R~\cite{Tolos:2016hhl}, 10~PKDD~\cite{Long:2003dn}, 11~TW99~\cite{Typel:1999yq}. Circles lie inside the band, crosses (2--6) outside. The dotted outline is the cold ($\shat=0$) region at the same density, shown for reference. The construction propagates the bound inward from the pQCD region along a fixed-$\shat$ ray.}
  \label{fig:overview}
\end{figure*}

\section{Setup: the model-independent bound at fixed entropy per baryon}
The model-independent construction of \cite{Komoltsev:2021jzg} constrains the EOS at intermediate densities between a low- and a high-density anchor based on causality, stability and thermodynamic consistency.
These two anchors should consist of a triple of thermodynamic quantities, $(v, \nB(v), p(v))$, as a function of some parameterizing thermodynamic variable $v$ on a given adiabat.
At zero temperature the natural parametrizing variable is the baryon chemical potential $\muB$, because the pressure is recovered by integrating $\mathrm{d}p=\nB\,\mathrm{d}\muB$ and $\muB$ increases monotonically with density.
Replacing the variable $\mu_B$ with the specific enthalpy
\begin{equation}
  h \equiv \frac{\varepsilon+p}{\nB} = \muB + T\,\shat ,
  \label{eq:h}
\end{equation}
extends these constraints directly to finite $\hat s$; the explicit constraints at finite $\hat s$ can be read from Ref.~\cite{Komoltsev:2021jzg} with simple replacement of the $\mu_B\rightarrow h$, and corresponding changes to the low- and high-density bounds. Here $\varepsilon$ is the energy density, and the second equality is the Euler relation ${\varepsilon+p=Ts+\muB\nB}$ for neutrino-transparent, charge-neutral matter in $\beta$ equilibrium. Differentiating at fixed entropy per baryon ($\mathrm{d}\shat=0$) gives $\mathrm{d}h=\mathrm{d}\muB+\shat\,\mathrm{d}T$, so the thermal terms recombine into the single, purely ``mechanical'' relation
\begin{equation}
  \mathrm{d}p = \nB\,\mathrm{d}h \qquad (\shat=\text{const}).
  \label{eq:dpdh}
\end{equation}
Thus on each adiabat $h$ plays the same parameterizing role as $\muB$ at $T=0$: it increases monotonically (since $\mathrm{d}p>0$) and integrates the pressure, so the entire cold construction carries over exactly under $\muB\to h$. 
Causality and stability place a lower bound on the slope $\mathrm{d}\nB/\mathrm{d}h$, while integrating Eq.~\eqref{eq:dpdh} from $h_\L$ to $h_\H$ must reproduce the pressure difference $\Delta p=p_\H-p_\L$. The analytic boundaries are given in Appendix~\ref{app:analytic}.

We use this finite-$T$ version of the construction of Ref.~\cite{Komoltsev:2021jzg} to constrain nuclear-physics models
at intermediate densities at varying $\shat$.
At a given $\nB$, we study whether the given nuclear models remain consistent with the given low- and high-density anchors as $\shat$ is increased.
We inspect as an example HS(DD2) and DD-MEX tables, a set of eleven RMF parametrizations, and two Bayesian posteriors of nucleonic RMF EOSs~\cite{Cartaxo:2025jpi}, so that the thermal effects on the bound are quantified both for individual models and statistically.

For the low-density anchor, we use the HS(DD2) isentrope at $\nB=0.176\fmc\,(\simeq1.1\,\nsat)$, taken from the CompOSE database~\cite{Typel:2013rza,CompOSECoreTeam:2022ddl} (entry HS(DD2)~\cite{Hempel:2009mc,Typel:2009sy}).
For the high-density anchor, we take a scale-averaged pQCD calculation as described in App.~\ref{app:pqcd} \cite{Gorda:2021kme,Gorda:2021gha}, tabulated on a regular $(\muB,T)$ grid whose cold edge lies at $\muB\simeq2.4~\mathrm{GeV}$ [See Fig.~\ref{fig:overview}(a)].
We map the grid onto the $(\nB,s)$ plane [Fig.~\ref{fig:overview}(b)] and read the anchor along the ray $s=\shat\,\nB$ at $\nB=6.5\fmc$. The anchor values are listed in Table~\ref{tab:anchors}.

\begin{table*}[tbp]
\begin{ruledtabular}
\begin{tabular}{ccccccc}
 & \multicolumn{3}{c}{low anchor (DD2)} & \multicolumn{3}{c}{high anchor (pQCD)}\\
\cline{2-4}\cline{5-7}
$\shat$ & $h_\L$ & $T_\L$ & $p_\L$ & $h_\H$ & $T_\H$ & $p_\H$\\
 & [GeV] & [MeV] & [$\frac{\mathrm{MeV}}{\mathrm{fm}^3}$] & [GeV] & [MeV] &
 [$\frac{\mathrm{MeV}}{\mathrm{fm}^3}$]\\
\hline
$0$ & $0.981$ & $\simeq0$ & $4.2$  & $2.620$ & $\simeq0$ & $3498$\\
$1$ & $0.995$ & $15.3$    & $5.3$  & $2.634$ & $12.3$    & $3534$\\
$2$ & $1.036$ & $32.2$    & $8.4$  & $2.675$ & $27.7$    & $3645$\\
$4$ & $1.198$ & $68.4$    & $19.1$ & $2.831$ & $73.8$    & $4088$\\
\end{tabular}
\end{ruledtabular}
\caption{Anchors of the finite-$\shat$ construction. The low-density anchor is the DD2 isentrope at fixed ${\nB^\L=0.176\fmc}$; the high-density anchor is the scale-averaged pQCD result at fixed $\nB^\H=6.5\fmc$. Listed are the specific enthalpy $h$, temperature $T$, and pressure $p$ at each anchor; the chemical potential follows from $\muB=h-T\shat$. At $\shat=0$, $h=\muB$ and the $T=0$ anchors are recovered.}
\label{tab:anchors}
\end{table*}

Figure~\ref{fig:overview}(c) shows the resulting allowed $(h,\nB)$ bands for $\shat=0,1,2,4$.
The bands overlap strongly at low $h$ but the upper (stiff) boundary recedes with $\shat$: at fixed enthalpy the allowed density window narrows as the matter is heated.
Projecting the band to a fixed baryon density yields an allowed region in the $(\varepsilon,p)$ plane; panel~(d) shows this region at $6\,\nsat$ for $\shat=2$, together with eleven relativistic mean-field (RMF) EOSs spanning the nonlinear and density-dependent families: FSU2R; the density-dependent set DD-ME1, DD-ME2, DD-MEX, DD-MEX1, DD-MEX2, TW99, PKDD, and DDV; and the chiral-EFT--informed parametrizations DDB~\cite{Malik:2022zol} and CHar~\cite{Char:2023fue}.
At this moderate entropy, five of the eleven models (DD-ME1, DD-ME2 and the three DD-MEX parametrizations) fall outside the model-independent band.
In fact, we find that as $\shat$ is increased from zero to four, the construction excludes progressively more models at this fixed density: for $\shat \le 1$, the construction excludes four of the eleven models (DD-ME2 and the DD-MEX family), while at $\shat=2$ it excludes five, and at $\shat = 4$ it excludes seven total (adding CHar and PKDD).
We also observe this trend at a lower density of $5\,\nsat$, where no model is excluded for $\shat \leq 2$ and five are excluded at $\shat=4$.
These counts refer to the central scale-averaged anchor; shifting the pQCD pressure across its scale-averaging uncertainty changes the cold count at $6\,\nsat$ between one and five without altering the trend with $\shat$.

\begin{figure}[t]
  \centering
  \includegraphics[width=\columnwidth]{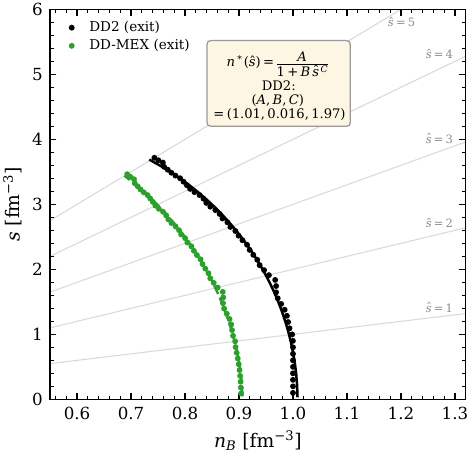}
  \caption{Exit arch on the $(\nB,s)$ plane for DD2 (black) and DD-MEX
  (green). Points are the computed exit densities $\nstar(\shat)$; curves are
  the universal fit, Eq.~\eqref{eq:fit}. Faint gray lines are radial $\shat$
  rays. Heating drives the exit to lower density.}
  \label{fig:arch}
\end{figure}

\section{Locating the breakdown of nuclear models as a function of entropy per baryon}

We now address a second possible question, namely, at which density does a given model EOS become inconsistent with pQCD as a function of $\shat$ when using it alone as the low-density anchor \cite{Gorda:2026rzm}?
For a given EOS we define the exit density $\nstar(\shat)$ as the highest baryon density at which the isentrope still lies inside the allowed $(h,\nB)$ band; beyond $\nstar$ the model is incompatible with pQCD plus causality and stability.
Figure~\ref{fig:arch} shows the exit locus $(\nstar,\,\shat\,\nstar)$ for DD2 and for the density-dependent model DD-MEX~\cite{Taninah:2019cku} (computed here with a finite-$T$ RMF solver). For cold matter both EOSs remain inside the band up to the end of their tabulation ($\nstar\simeq1.0$ and $0.9\fmc$, respectively), but as $\shat$ increases the exit density falls steadily, tracing a characteristic ``arc'' in the $(\nB,s)$ plane. The locus is captured to better than a percent by the three-parameter form
\begin{equation}
  \nstar(\shat) = \frac{A}{1+B\,\shat^{\,C}},
  \label{eq:fit}
\end{equation}
with $(A,B,C)=(1.01,0.016,1.97)$ for DD2 and $(0.91,0.016,1.85)$ for DD-MEX. The near-universality of $B$ and $C$ across these two structurally different RMF families suggests that the shape of the thermal weakening is set by the pQCD/causality geometry, while the normalization $A$ carries the low-density nuclear physics.

\begin{figure*}[t]
  \centering
  \includegraphics[width=\textwidth]{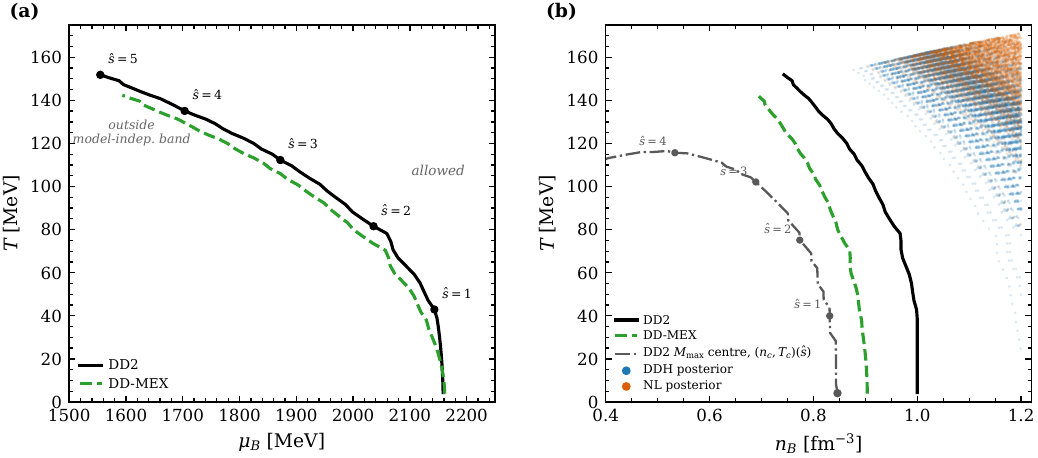}
  \caption{The exit boundary in the QCD phase diagram. \textbf{(a)}  $(\muB,T)$ plane: DD2 (solid) and DD-MEX (dashed) exit loci, labeled by $\shat$; matter to the high-$\muB$ side is still allowed. \textbf{(b)} $(\nB,T)$ plane: the same loci with the NL/DDH posterior exit clouds; the dash-dotted curve is the central density and temperature of the maximum-mass DD2 star computed on the same isentropes, from the cold star ($\nB^c=0.85\fmc$) to $\shat=4.6$. The boundary sweeps through the merger/PNS regime.}
  \label{fig:phase}
\end{figure*}

By $\shat=5$ the DD2 exit density has dropped to ${\nstar\simeq0.74\fmc}$, below the central density of the maximum-mass cold DD2 star ($\nB^c\simeq0.85\fmc$).
The comparison with a hot star is less favourable: heating inflates the star faster than it lowers $\nstar$, so the central density of the maximum-mass configuration on the same isentrope falls to $0.77$, $0.52$ and $0.37\fmc$ at $\shat=2$, $4$ and $4.6$ [dash-dotted curve in Fig.~\ref{fig:phase}(b)], and the exit locus stays outside every isentropic DD2 star.
The bound therefore acts on the finite-$T$ EOS at densities that isentropic stars do not reach; whether it is approached in a merger remnant, where hot matter sits at densities set by the cold core rather than by hydrostatic equilibrium on a single isentrope, has to be judged against simulation data.

\begin{figure}[t]
  \centering
  \includegraphics[width=\columnwidth]{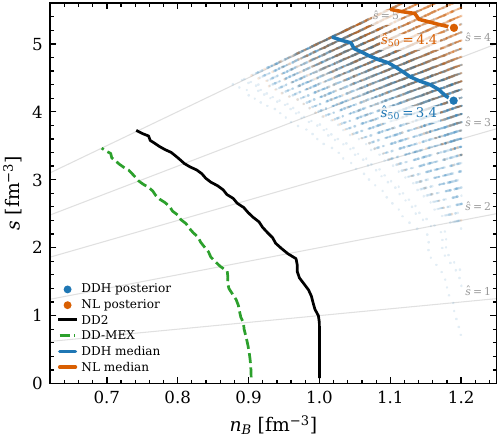}
  \caption{Constraining power over Bayesian RMF posteriors. Exit clouds on
  the $(\nB,s)$ plane for the NL (orange) and DDH (blue) posteriors, with the
  DD2 and DD-MEX exit curves for reference; faint gray lines are radial $\shat$
  rays. Thick lines are the posterior median of the exit density, taken over
  the full posterior with samples that have not left the band treated as
  censored above the search window ($\nB>1.2\fmc$); a median exists only on
  rays along which at least half of the posterior has exited, so each line
  begins at that ray (filled circle), $\shat_{50}\simeq3.4$ (DDH) and
  $\simeq4.4$ (NL).}
  \label{fig:violation}
\end{figure}

To assess the statistical reach of the bound we propagate two Bayesian posteriors of nucleonic RMF EOSs, a density-dependent (DDH, Typel99) set and a nonlinear (NL) set, through the construction, computing $\nstar(\shat)$ for $500$ samples of each. Figure~\ref{fig:violation} shows the resulting exit
clouds: the DDH posterior reaches the band at systematically lower $\shat$ than
the NL posterior, reflecting its softer high-density behavior. The thick
lines give the posterior median of the exit density at fixed $\shat$, taken
over the full posterior with samples that have not yet left the band treated as
censored above the search window; a median therefore exists only on rays along
which at least half of the posterior has already exited, and each line begins
at that ray, $\shat_{50}\simeq3.4$ (DDH) and $\shat_{50}\simeq4.4$ (NL). By
$\shat=5$ essentially all DDH models and $\sim90\%$ of NL models are excluded.
Consistency with the cold pQCD bound does not guarantee consistency along finite-entropy adiabats. For the two Bayesian posteriors studied here, the fraction of excluded models grows strongly with entropy per baryon.

Finally, Fig.~\ref{fig:phase} recasts the exit locus into the QCD phase diagram. Using Eq.~\eqref{eq:h} to convert each exit point to $\muB=h-T\shat$, the DD2 and DD-MEX exit boundaries sweep from $(\muB,T)\simeq(2.16~\mathrm{GeV},\,0)$ at $\shat\to0$ to $(1.55~\mathrm{GeV},\,152~\mathrm{MeV})$ at $\shat=5$ [panel~(a)]. In the $(\nB,T)$ plane [panel~(b)] the DD2 boundary passes the cold maximum-mass central density $\nB^c=0.85\fmc$ at $T\simeq125~\mathrm{MeV}$ ($\shat\simeq3.5$), but remains at higher density than the centre of the maximum-mass star on every isentrope (dash-dotted curve). This is the temperature--density region reached transiently in the post-merger remnant and in the hot PNS, so that the bound constrains the physical states sampled by current and future multimessenger observations.

\section{Conclusions}

At finite temperature a fully model-agnostic inference of the kind available for cold matter is out of reach: the astrophysical data that would constrain hot dense matter are sparse and indirect, and simulations will continue to rely on nuclear-physics models for the EOS tables that close their equations. Those models should, however, be built with every available constraint, and the construction presented here adds one that has so far been missing: a first-principles, model-independent condition connecting any finite-temperature EOS to pQCD along adiabats. Its content is compact: the exit density $\nstar(\shat)$ and its three-parameter law, Eq.~\eqref{eq:fit}, the tightening of the $(\varepsilon,p)$ region with $\shat$, and the half-exclusion entropies $\shat_{50}$ of Fig.~\ref{fig:violation} provide model-independent targets against which existing tables can be checked and future ones calibrated. For the models tested here, the bound is already active for cold matter at $6\,\nsat$ and becomes increasingly restrictive with entropy per baryon; approximately half of the DDH and NL posterior samples are excluded within the density search window at $\shat\simeq3.4$ and $4.4$, respectively. We intend to fold the bound into Bayesian finite-$T$ EOS inference as a likelihood term, to replace the low-density anchor by a chiral-EFT one so that both ends of the construction are first-principles, and to extend the construction to fixed lepton fraction for supernova and proto-neutron-star applications.

\begin{acknowledgments}
The authors are listed alphabetically. A.K. was supported by the Research Council of Norway through the FRIPRO programme (CoreQCD, project number 361873) and in part by grant NSF PHY-2309135 to the Kavli Institute for Theoretical Physics (KITP). T.M.\ would like to acknowledge the support of FCT (Funda\c{c}\~{a}o para a Ci\^{e}ncia e a Tecnologia, I.P., Portugal) under project UID/04564/2025, identified by DOI 10.54499/UIDB/04564/2025. T.M.\ would also like to thank the FCT Mobility grant (Reference: FCT/Mobility/1381803446/2024-25) and the University of Stavanger, Department of Mathematics and Physics, for the local hospitality during the visit for this work.
\end{acknowledgments}

\appendix

\section{\texorpdfstring{Analytic finite-$\shat$ construction}{Analytic construction at finite entropy}}
\label{app:analytic}
The finite-temperature bound used in the main text is built numerically, but it also admits closed-form boundaries. Here we derive them and show that they reproduce the numerical construction exactly.

\subsection{Enthalpy-plane formulation}
The Euler relation $\varepsilon+p=Ts+\muB\nB$ gives, per baryon, the specific enthalpy
\begin{equation}
  h \equiv \frac{\varepsilon+p}{\nB} = \muB + T\,\shat ,
  \label{eq:a-h}
\end{equation}
with $\shat=s/\nB$. Differentiating along an adiabat ($\mathrm{d}\shat=0$) and combining with the Gibbs--Duhem relation
$\mathrm{d}p=s\,\mathrm{d}T+\nB\,\mathrm{d}\muB$,
\begin{equation}
  \mathrm{d}h=\mathrm{d}\muB+\shat\,\mathrm{d}T
  \;\Longrightarrow\;
  \nB\,\mathrm{d}h=\nB\,\mathrm{d}\muB+s\,\mathrm{d}T=\mathrm{d}p .
  \label{eq:a-dpdh}
\end{equation}
Similarly, $\varepsilon=h\nB-p$ gives $\mathrm{d}\varepsilon=h\,\mathrm{d}\nB$ at fixed $\shat$, so that
\begin{equation}
  c_s^2=\left(\frac{\mathrm{d}p}{\mathrm{d}\varepsilon}\right)_{\shat}
  =\left(\frac{\mathrm{d}\ln h}{\mathrm{d}\ln \nB}\right)_{\shat}.
  \label{eq:a-cs2}
\end{equation}
Equations~\eqref{eq:a-dpdh}--\eqref{eq:a-cs2} are formally identical to the $T=0$ relations $\mathrm{d}p=\nB\,\mathrm{d}\muB$, $c_s^2=(\mathrm{d}\ln\muB/\mathrm{d}\ln \nB)$ under $\muB\to h$. The Komoltsev--Kurkela construction therefore applies verbatim on each adiabat with the chemical potential replaced by the specific enthalpy; at $\shat=0$, $h=\muB$ and the original $T=0$ bound is recovered.

\begin{figure*}[t]
  \centering
  \includegraphics[width=\textwidth]{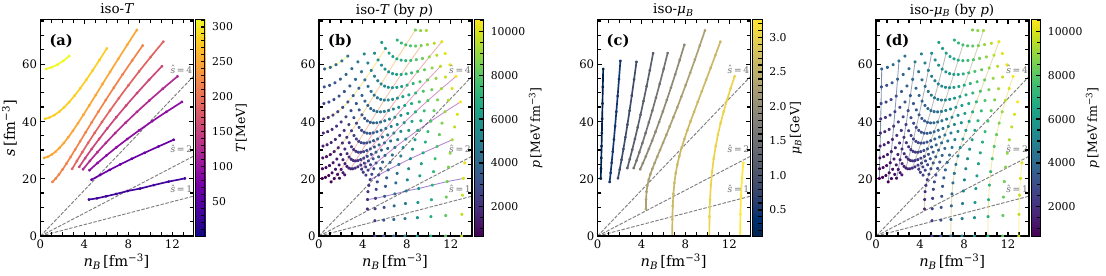}
  \caption{Scale-averaged pQCD data on the $(\nB,s)$ plane. \textbf{(a)} Iso-$T$ lines colored by $T$. \textbf{(b)} Iso-$T$ lines with markers colored by pressure $p$. \textbf{(c)} Iso-$\muB$ lines colored by $\muB$. \textbf{(d)} Iso-$\muB$ lines with markers colored by $p$. Dashed lines are the radial $\shat=s/\nB$ rays, along which the high-density anchor is read at
  $\nB=6.5\fmc$.}
  \label{fig:pqcdiso}
\end{figure*}

\subsection{Closed-form band boundaries}
Given the low anchor $(h_\L,\nB^\L,p_\L)$ and high anchor $(h_\H,\nB^\H,p_\H)$, define $C_\L=\nB^\L/h_\L$, $C_\H=\nB^\H/h_\H$, and $\Delta p=p_\H-p_\L$. Causality and stability bound the slope $C_\L\le\mathrm{d}\nB/\mathrm{d}h$ via the $c_s^2=1$ lines $\nB=C_{L,H}\,h$ through the origin, while the pressure constraint $\int_{h_\L}^{h_\H}\nB\,\mathrm{d}h=\Delta p$ [Eq.~\eqref{eq:a-dpdh}] fixes the enclosed area. The extremal isentropes are piecewise causal segments, and the elementary integrals ($\int C x\,\mathrm{d}x=\tfrac{C}{2}\Delta x^2$) yield the band boundaries in closed form,
\begin{align}
  \nB^{\max}(h)&=\min\!\left[\frac{2h\big(\Delta p-\tfrac{C_\L}{2}(h^2-h_\L^2)\big)}
       {h_\H^2-h^2},\; C_\H\,h\right],\label{eq:a-nmax}\\[4pt]
  \nB^{\min}(h)&=\max\!\left[\frac{2h\big(\Delta p-\tfrac{C_\H}{2}(h_\H^2-h^2)\big)}
       {h^2-h_\L^2},\; C_\L\,h\right],\label{eq:a-nmin}
\end{align}
where the caps are the causal cone. The rational branch and its cap meet at
\begin{equation}
  h_c=\sqrt{\frac{h_\L h_\H\,(h_\H\nB^\H-h_\L\nB^\L-2\Delta p)}{h_\L\nB^\H-h_\H\nB^\L}} .
  \label{eq:a-hc}
\end{equation}
Equations~\eqref{eq:a-nmax}--\eqref{eq:a-hc} are Eqs.~(6),(7),(9) of Ref.~\cite{Komoltsev:2021jzg} under $\muB\to h$. Projecting at fixed $\nB$ via $\varepsilon=h\nB-p$ gives the $(\varepsilon,p)$ band
\begin{align}
  p_{\min}(h)&=\max\!\left[p_\L+\tfrac{C_\L}{2}(h^2-h_\L^2),\;
      p_\H-\tfrac{C_\H}{2}(h_\H^2-h^2)\right],\\
  p_{\max}(h)&=\min\!\left[p_\L+\tfrac{\nB}{2h}(h^2-h_\L^2),\;
      p_\H-\tfrac{\nB}{2h}(h_\H^2-h^2)\right],
\end{align}
\begin{equation}
  \begin{aligned}
    \varepsilon_{\min}(h)&=h\,\nB-p_{\max}(h),\\
    \varepsilon_{\max}(h)&=h\,\nB-p_{\min}(h),
  \end{aligned}
\end{equation}
valid over the window $\nB^{\min}(h)\le \nB\le \nB^{\max}(h)$. These closed forms reproduce the numerical envelope used in the main text, and at $\shat=0$ they coincide with the published $T=0$ bound \cite{Komoltsev:2021jzg}.

\section{Scale-averaged pQCD input}
\label{app:pqcd}

The scale-averaged pQCD data set is constructed as follows.
We use the finite-$T$ results of Ref.~\cite{Gorda:2021gha}, which adds strange-quark mass effects to the  massless $O(\alpha_s^{5/2})$ results of Ref.~\cite{Kurkela:2016was}.
This allows us to compute all of the thermodynamic variables in the perturbative regime, with an uncertainty quantified by the choice of $\overline{\text{MS}}$ renormalization scale $\bar\Lambda$.

To set the scale, we use the following argument.
As has long been known~\cite{Kajantie:2002wa}, the high-$T$, zero-$\muB$ pQCD result can be matched very well to lattice QCD results for a particular value of $\bar\Lambda$ .
For this reason, we choose to use the following parametrization for the renormalization scale as a function of $T$ and $\muB$
\begin{equation}
    \bar\Lambda = \sqrt{(0.722 \times 4 \pi T X_\mathrm{lat})^2 + (2 \muB X / 3)^2},
\end{equation}
which separately parametrizes the $T = 0$ and $\muB = 0$ pQCD uncertainties.
We find that $X_\mathrm{lat} \in [0.712, 0.812]$ reproduces the lattice-QCD uncertainties $s$ and $p$ reported in Ref.~\cite{Borsanyi:2013bia}, 
while we use $X \in [0.5, 2.0]$ as usual \cite{Kurkela:2009gj}.
In our calculation, we perform a log-uniform averaging in $X$ and $X_\mathrm{lat}$ by choosing to relate $X_\mathrm{lat}$ to $X$ as
\begin{equation}
    X_\mathrm{lat} = (X-0.5)/15 + 0.712,
\end{equation}
and perform a log-uniform average over $X \in [0.5, 2.0]$.
In this way, the errors in the pQCD result will grow from the lattice uncertainties at $\muB = 0$ towards the usual scale-variation uncertainties as one moves towards the $T = 0$ axis.

Fig.~\ref{fig:pqcdiso} shows the scale-averaged pQCD data set used for the high-density anchor, mapped from its native regular $(\muB,T)$ grid onto the $(\nB,s)$ plane. Panels (a,b) display lines of constant temperature and panels (c,d) lines of constant baryon chemical potential; markers in (b,d) are colored by pressure. The data form a smooth curvilinear mesh, and the high-density anchor at each $\shat$ is obtained by evaluating the linear interpolant of this mesh at $\nB=6.5\fmc$ along the radial ray $s=\shat\,\nB$ (dashed). The mesh is densest in the anchor region $\nB\gtrsim6\fmc$. The image of the regular $(\muB,T)$ grid is a curvilinear mesh that is densest exactly where the anchor is read, and we evaluate it by linear interpolation. For all $\shat\le5$ the anchor point $(\nB,s)=(6.5,\,6.5\shat)\fmc$ lies inside the tabulated region, but with increasing $\shat$ it approaches the low-$\muB$ edge of the grid: the nearest grid node moves from $\muB=2.66~\mathrm{GeV}$ for $\shat\le2$ to $\muB=2.41~\mathrm{GeV}$, $T=114~\mathrm{MeV}$ at $\shat=5$, where the scale-averaged result carries its largest uncertainty. Across the scale-averaging band the anchor pressure changes by about $\pm20\%$ ($3.50\pm0.74~\mathrm{GeV}\fmc$ at $\shat=0$); all exit densities quoted in the main text use the central value.

\bibliographystyle{apsrev4-2}
\bibliography{references}

\end{document}